# The Sun and the Moon a Riddle in the Sky

Uri Lachish

urila@zahav.net.il

**Abstract:** The uniform image of the full moon is well known from the beginning of history. In the last decades there are photos with similar configuration of the earth observed from the moon and from space, as well as of all the planets and their moons. The photos are all nearly uniform. Such images are considered non-Lambertian since they do not comply with Lambert Cosine Law of light scattering. Theories of the uniformity deal only with the moon. Apart from being not persuasive for the moon case, they are not applicable to most other cases. There are thousands of thousands of similar photos, but there is not a single true photo that does comply with Lambert cosine Law. Photos that do comply with the law are all "rendered", that is, at least partly simulated. A calculation based on fundamental principles is presented to clarify the uniformity in all cases regardless of the observed object and its surface properties. The uniformity is a direct outcome of a single sunlight scattering of the observed object, and there is no need of further assumptions or models to justify it.

The sun image is uniform except some marginal blur at its periphery. The full-moon image shows some details of mountains and dry lakes, but apart from that the moon image is uniform as that of the sun.

The following two images in fig.-1 show the sun and the full-moon:

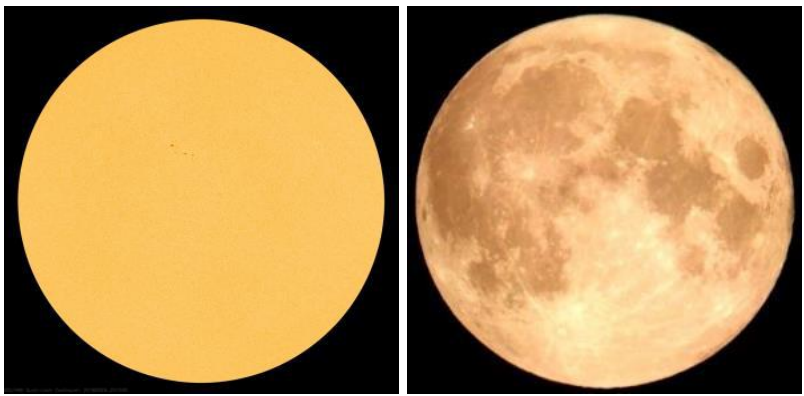

Fig.-1: The sun[1] and the full moon[2] images. Both have a uniform brightness distribution.

The sun is a Lambertian light source, that is, each point on its surface scatters light equally in all directions. The moon scatters the sun light that falls on its surface, and according to Lambert's cosine law it will be most bright at its face center, and the light intensity will fade to zero when moving toward its periphery.

However, as fig.-1 shows, and as is seen by the eye each month, the full-moon image is uniform, and theories of non-Lambertian light scattering of the moon surface have been suggested in order to explain the observation [3, 4]. These sources also made a thorough literature search of the phenomenon.

Photos of all the planets and the planet moons, taken from space with the sun at the back of the observer, show similar uniformity [5]. Mars is similar to the moon, but Venus is covered with heavy clouds and the sunlight is scattered from a gas phase.

Fig.-2 shows an image of the earth taken from the moon [6]:

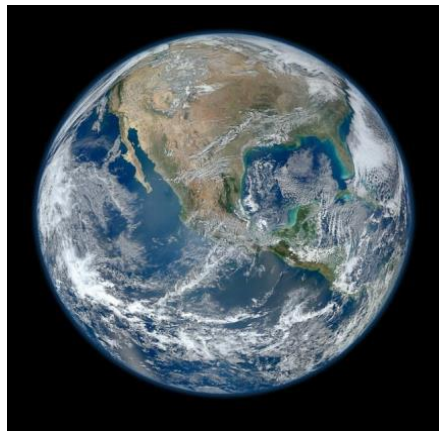

Fig.-2: The earth image taken from the moon.

The image includes large areas of solid land, liquid water and gas clouds, and the uniformity is observed with each of them. It is, therefore, unlikely that the observed uniformity is the outcome of some property of the scattering medium. Models of light scattering that correspond to this case, deal only with solid surfaces. Explanation or models that work for scattering from the three phases are difficult to find.

In addition to the full-moon, there are many images of spheres and other curved surfaces with similar directional illumination, but true photos of any of them, that obeys Lambert's cosine law, are again difficult to find. It seems that all the photos that obey the law are at least partly simulated.

When an electromagnetic wave advances in a medium it polarizes its matter, and each polarized dipole is a source of electromagnetic wave. The effect of all the dipoles is calculated, for example, by vector addition of their electric fields. Each dipole has a different phase that depends on its position, and if the material is uniform these fields cancel each other in all the directions except the forward direction, where the effect is refraction.

Scattering comes from non-uniform fluctuations of material density, and then scattered waves may advance in any direction. The intensity of scattered light is the sum of the intensities of the scattering centers and it does not depend on the phases of the scattering dipoles.

The purpose of this discussion is to indicate that the uniformity of the moon image is a direct outcome of single light scattering. Further mechanisms or models are not necessary. The moon surface is Lambertian after all in the sense that the scattering is random, although it does not obey Lambert cosine law.

Consider an imaginary line between the sun and the earth, or between the sun and the moon, and a unit cross section area *a* on a plane perpendicular to that line.

Fig.-3 shows an observer on the earth looking directly at the sun (left), and looking at night at the full moon, while the sun is behind his back (right).

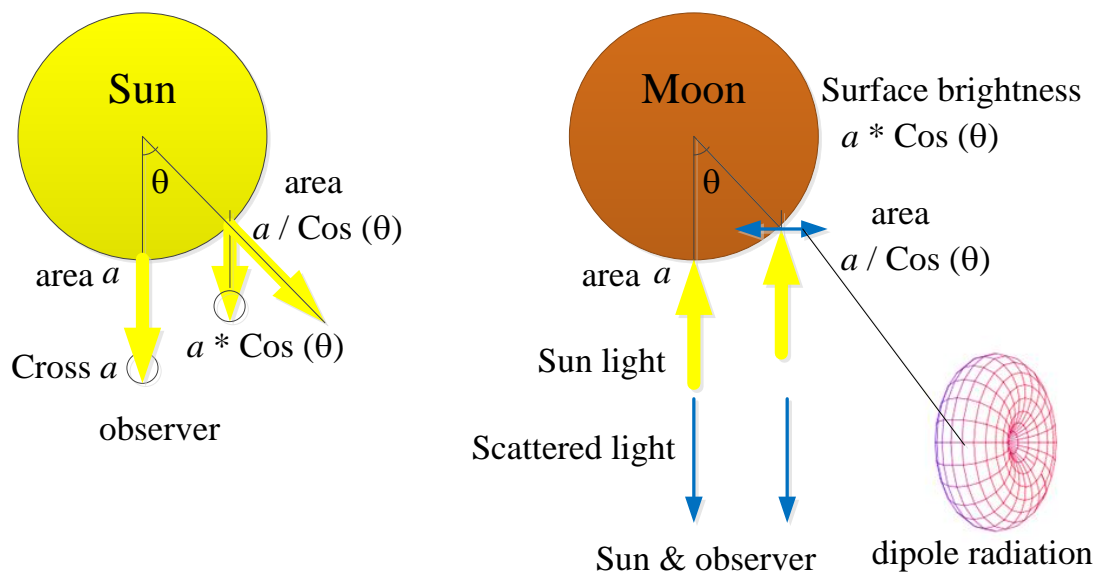

Fig.-3: Relative configurations of the sun, the moon and an observer on the earth.
   (1) Sun surface area $a / \cos(\theta)$ is observed through a cross section area *a*.
       Sunlight $a * \cos(\theta)$ is directed to an observer on the earth.

   (2) Sunlight brightness $a * \cos(\theta)$ on the moon illuminates the area $a / \cos(\theta)$ through *a*. The sun light polarizes the moon surface perpendicular to the light direction. The moon dipoles scatter maximum light back to the sun, and back to an observer on the earth.

$\theta$ is the angle between a line from the sun center to an observer on the earth, and a line between the sun center and any point on its surface.

The sun is a Lambertian source of light emitted uniformly from its surface, fig.-3 (left). That is, each point on the surface emits equal light in all the directions of the solid angle $2\pi$, and the mean equivalent light is perpendicular to the sun surface. $\theta$ defines the position of any point on the sun surface. $\theta$ similarly defines any position on the moon surface.

Consider a line between the sun and the earth, or between the sun and the moon, and a unit cross section area *a* on a plane perpendicular to that line.

The area on the sun surface observed through the cross section area *a* will be:

$$a / \cos(\theta), \qquad (1)$$

Thus, it is equal to *a* at the sun center, and it increases toward the periphery. The mean light emitted from that area is perpendicular to the sun surface, and its component seen by an observer on earth will be proportional to $\cos(\theta)$.

Therefore the sun-light observed through the cross section *a* will be proportional to:

$$Is * (a / \cos(\theta)) * \cos(\theta). \qquad (2)$$

Where *Is* is the sunlight density at its surface. Hence, the light toward the observer does not depend on $\cos(\theta)$ and the sun brightness observed from earth will be uniform [7-8].

The case of the moon, fig.-3 (right), is discussed in a similar manner. However, moonlight is a scattered sunlight, and the sunlight on the moon surface is $Im * \cos(\theta)$, where *Im* is the sunlight density at the moon face center.

Repeating the same argument, the full moonlight would be proportional to:

$$Im * \cos(\theta) * (a / \cos(\theta) * \cos(\theta)). \qquad (3)$$

The left $\cos(\theta)$ in the Eq. (3) comes from the fading light intensity on the moon surface. So, by Eq. (3), the moon image should also fade to zero from its face center toward its periphery. This is not the case as observed in fig.-1 (right), or by looking at the full moon.

It is beneficial to go a little deeper into the scattering mechanism. Sunlight that falls on the moon, induces dipole charge oscillations on its face. Each such a dipole is a source of light, and the overall scattered light is the combined effect of all the dipoles.

Each dipole oscillates in the electric field direction of the coming sunlight, that is, in a plane perpendicular to the light direction, and the maximal emission of a dipole is perpendicular to its direction of oscillation, that is, back to the sun.

Therefore, the mean equivalent moonlight will also be directed back to the sun, and in full moon also to an observer on the earth. It will not be perpendicular to the moon surface, as was the case with the sun surface. There is no need for the right Cos($\theta$) of Eq. (2) in Eq. (3).

Repeating the argument with fading sunlight on the moon surface, the back scattered moonlight will be proportional then to:

$$(a / \cos(\theta)) * \cos(\theta), \qquad (4)$$

Observer on earth will see a uniform full moon.

In summary, the mean back scattering from the full moon is directed back to the sun because the scattering dipoles on the moon oscillate in a plane perpendicular to the coming sunlight. Any calculation that does not take the direction of the polarizing dipoles into account will not be correct.

Back scattering yields a uniform moon image, as well as the earth image and the images of all the planets and their moons.

Back-scattering from any surface will not depend on the surface inclination angle to the coming light, and a curved surface will look uniform. It is, therefore, not surprising that true images, that obey Lambert's cosine law, are difficult to find.

In the calculation of Lambert's cosine law there is a hidden assumption that the scattering dipoles oscillate in all random directions in space. This is not the case with the moon, and probably with other examples. In the case of single scattering, the scattered light conserves the polarization plane of the coming light and lead to uniform surface image. In multiple scattering the scattered light loose this polarization, then it should obey Lambert cosine law according to eq. (3).

However, as a process involves more and more scattering steps, the probability for it will come down, and the first single scattering will be dominant.

The full moon image indicates that single scattering mechanism is responsible for the uniformity. In some cases multiple scattering may add some non-uniformity.

References

[1] Sun image https://sdo.gsfc.nasa.gov/assets/img/latest/latest_1024_HMIIF.jpg

[2] Moon image http://www.newsfour.ie/wp-content/uploads/2018/02/moon-1.jpg

[3] Z.K. Kopal, *An Itroduction to the Study of the Moon*, (Springer, 1966), p. 330

http://pdf.to/bookinfo/an-introduction-to-the-study-of-the-moon.pdf/

[4] Michael K. Shepard, Introduction to Planetary Photometry, Contemporary Physics, 59(1):1-1 · October 2017, (Cambridge, 2017)

[5] True-Color Photos of All the Planets: https://owlcation.com/stem/True-Color-Photos-of-All-the-Planets

[6] Earth image: https://qz.com/458826/behold-nasas-first-image-in-decades-of-the-whole-earth/

[7] W. J. Smith, *Modern Optical engineering*, 3rd Ed., (McGrow-Hill, NY, 2000), pp. 221-222.

[8] F. L. Pedrotti, L. S. Pedrotti, *Introduction to Optics*, 2nd Ed. (Prentic-Hall, NJ, 1993), pp. 11-12.

version 1:  http://urila.tripod.com/Moon ver-1.pdf

present version:  http://urila.tripod.com/Moon.pdf

Poster: https://www.researchgate.net/publication/339435329_The_Sun_and_the_Moon_a_Riddle_in_the_Sky

DOI: 10.13140/RG.2.2.17485.08166

*Think that this page is correct? Please pass it on to others.*